\documentclass[10pt,english]{article}
\usepackage{ae,aecompl}
\usepackage[T1]{fontenc}
\usepackage[latin9]{inputenc}
\usepackage{babel}
\usepackage{verbatim}
\usepackage{bm}
\usepackage{amsmath}
\usepackage{amssymb}
\usepackage{graphicx}
\usepackage{esint}
\PassOptionsToPackage{normalem}{ulem}
\usepackage{ulem}
\usepackage[unicode=true,pdfusetitle,
 bookmarks=true,bookmarksnumbered=false,bookmarksopen=false,
 breaklinks=false,pdfborder={0 0 1},backref=false,colorlinks=false]
 {hyperref}

\makeatletter
\usepackage{babel}
\usepackage[font=small,labelfont=bf]{caption}
\@ifundefined{definecolor}{\@ifundefined{definecolor}
 {\usepackage{color}}{}
}{}
\@ifundefined{definecolor}{\@ifundefined{definecolor}
 {\usepackage{color}}{}
}{}
\@ifundefined{definecolor}{\@ifundefined{definecolor}
 {\@ifundefined{definecolor}
 {\@ifundefined{definecolor}
 {\usepackage{color}}{}
}{}
}{}
}{}
\@ifundefined{definecolor}{\@ifundefined{definecolor}
 {\@ifundefined{definecolor}
 {\@ifundefined{definecolor}
 {\@ifundefined{definecolor}
 {\usepackage{color}}{}
}{}
}{}
}{}
}{}
\@ifundefined{definecolor}{\@ifundefined{definecolor}
 {\@ifundefined{definecolor}
 {\@ifundefined{definecolor}
 {\@ifundefined{definecolor}
 {\@ifundefined{definecolor}
 {\usepackage{color}}{}
}{}
}{}
}{}
}{}
}{}
\@ifundefined{definecolor}{\@ifundefined{definecolor}
 {\@ifundefined{definecolor}
 {\@ifundefined{definecolor}
 {\@ifundefined{definecolor}
 {\@ifundefined{definecolor}
 {\@ifundefined{definecolor}
 {\usepackage{color}}{}
}{}
}{}
}{}
}{}
}{}
}{}
\@ifundefined{definecolor}{\@ifundefined{definecolor}
 {\@ifundefined{definecolor}
 {\@ifundefined{definecolor}
 {\@ifundefined{definecolor}
 {\@ifundefined{definecolor}
 {\@ifundefined{definecolor}
 {\@ifundefined{definecolor}
 {\usepackage{color}}{}
}{}
}{}
}{}
}{}
}{}
}{}
}{}
\@ifundefined{definecolor}{\@ifundefined{definecolor}
 {\@ifundefined{definecolor}
 {\@ifundefined{definecolor}
 {\@ifundefined{definecolor}
 {\@ifundefined{definecolor}
 {\@ifundefined{definecolor}
 {\@ifundefined{definecolor}
 {\@ifundefined{definecolor}
 {\usepackage{color}}{}
}{}
}{}
}{}
}{}
}{}
}{}
}{}
}{}
\@ifundefined{definecolor}{\@ifundefined{definecolor}
 {\@ifundefined{definecolor}
 {\@ifundefined{definecolor}
 {\@ifundefined{definecolor}
 {\@ifundefined{definecolor}
 {\@ifundefined{definecolor}
 {\@ifundefined{definecolor}
 {\@ifundefined{definecolor}
 {\@ifundefined{definecolor}
 {\usepackage{color}}{}
}{}
}{}
}{}
}{}
}{}
}{}
}{}
}{}
}{}
\@ifundefined{definecolor}{\@ifundefined{definecolor}
 {\@ifundefined{definecolor}
 {\@ifundefined{definecolor}
 {\@ifundefined{definecolor}
 {\@ifundefined{definecolor}
 {\@ifundefined{definecolor}
 {\@ifundefined{definecolor}
 {\@ifundefined{definecolor}
 {\@ifundefined{definecolor}
 {\@ifundefined{definecolor}
 {\usepackage{color}}{}
}{}
}{}
}{}
}{}
}{}
}{}
}{}
}{}
}{}
}{}
\@ifundefined{definecolor}{\@ifundefined{definecolor}
 {\@ifundefined{definecolor}
 {\@ifundefined{definecolor}
 {\@ifundefined{definecolor}
 {\@ifundefined{definecolor}
 {\@ifundefined{definecolor}
 {\@ifundefined{definecolor}
 {\@ifundefined{definecolor}
 {\@ifundefined{definecolor}
 {\@ifundefined{definecolor}
 {\@ifundefined{definecolor}
 {\usepackage{color}}{}
}{}
}{}
}{}
}{}
}{}
}{}
}{}
}{}
}{}
}{}
}{}
\@ifundefined{definecolor}{\@ifundefined{definecolor}
 {\@ifundefined{definecolor}
 {\@ifundefined{definecolor}
 {\@ifundefined{definecolor}
 {\@ifundefined{definecolor}
 {\@ifundefined{definecolor}
 {\@ifundefined{definecolor}
 {\@ifundefined{definecolor}
 {\@ifundefined{definecolor}
 {\@ifundefined{definecolor}
 {\@ifundefined{definecolor}
 {\usepackage{color}}{}
}{}
}{}
}{}
}{}
}{}
}{}
}{}
}{}
}{}
}{}
}{}
\@ifundefined{definecolor}{\@ifundefined{definecolor}
 {\@ifundefined{definecolor}
 {\@ifundefined{definecolor}
 {\@ifundefined{definecolor}
 {\@ifundefined{definecolor}
 {\@ifundefined{definecolor}
 {\@ifundefined{definecolor}
 {\@ifundefined{definecolor}
 {\@ifundefined{definecolor}
 {\@ifundefined{definecolor}
 {\@ifundefined{definecolor}
 {\@ifundefined{definecolor}
 {\@ifundefined{definecolor}
 {\usepackage{color}}{}
}{}
}{}
}{}
}{}
}{}
}{}
}{}
}{}
}{}
}{}
}{}
}{}
}{}
\@ifundefined{definecolor}{\@ifundefined{definecolor}
 {\@ifundefined{definecolor}
 {\@ifundefined{definecolor}
 {\@ifundefined{definecolor}
 {\@ifundefined{definecolor}
 {\@ifundefined{definecolor}
 {\@ifundefined{definecolor}
 {\@ifundefined{definecolor}
 {\@ifundefined{definecolor}
 {\@ifundefined{definecolor}
 {\@ifundefined{definecolor}
 {\@ifundefined{definecolor}
 {\@ifundefined{definecolor}
 {\@ifundefined{definecolor}
 {\usepackage{color}}{}
}{}
}{}
}{}
}{}
}{}
}{}
}{}
}{}
}{}
}{}
}{}
}{}
}{}
}{}
\usepackage{bm}

\@ifundefined{definecolor}{\@ifundefined{definecolor}
 {\@ifundefined{definecolor}
 {\@ifundefined{definecolor}
 {\@ifundefined{definecolor}
 {\@ifundefined{definecolor}
 {\@ifundefined{definecolor}
 {\@ifundefined{definecolor}
 {\@ifundefined{definecolor}
 {\@ifundefined{definecolor}
 {\@ifundefined{definecolor}
 {\@ifundefined{definecolor}
 {\@ifundefined{definecolor}
 {\@ifundefined{definecolor}
 {\@ifundefined{definecolor}
 {\@ifundefined{definecolor}
 {\usepackage{color}}{}
}{}
}{}
}{}
}{}
}{}
}{}
}{}
}{}
}{}
}{}
}{}
}{}
}{}
}{}
}{}

\usepackage{bm}

\@ifundefined{definecolor}{\@ifundefined{definecolor}
 {\@ifundefined{definecolor}
 {\@ifundefined{definecolor}
 {\@ifundefined{definecolor}
 {\@ifundefined{definecolor}
 {\@ifundefined{definecolor}
 {\@ifundefined{definecolor}
 {\@ifundefined{definecolor}
 {\@ifundefined{definecolor}
 {\@ifundefined{definecolor}
 {\@ifundefined{definecolor}
 {\@ifundefined{definecolor}
 {\@ifundefined{definecolor}
 {\@ifundefined{definecolor}
 {\@ifundefined{definecolor}
 {\@ifundefined{definecolor}
 {\usepackage{color}}{}
}{}
}{}
}{}
}{}
}{}
}{}
}{}
}{}
}{}
}{}
}{}
}{}
}{}
}{}
}{}
}{}

\usepackage{amsfonts}

\usepackage{epsfig}

\usepackage{latexsym}

\def\lesssim{\mathrel{\hbox{\rlap{\hbox{\lower4pt\hbox{$\sim$}}}\hbox{$<$}}}}
\def\gtrsim{\mathrel{\hbox{\rlap{\hbox{\lower4pt\hbox{$\sim$}}}\hbox{$>$}}}}

\usepackage{aecompl}
\usepackage{cite}
\date{}

\makeatother

\begin{document}
\title{\textbf{{Reply to the comment on ``Frame-dragging: meaning, myths,
and misconceptions'' by A. Deriglazov }}}
\author{L. Filipe O. Costa$^{1}$\thanks{lfilipecosta@tecnico.ulisboa.pt},
José Natário$^{1}$\thanks{jnatar@math.ist.utl.pt} \\
 {\small{}{}{} {\em $^{1}$CAMGSD --- Departamento de Matemática,
Instituto Superior Técnico,} }\\
 {\small{}{}{} {\em Universidade de Lisboa, 1049-001, Lisboa,
Portugal}}}
\maketitle
\begin{abstract}
It has been claimed in {[}\href{https://arxiv.org/abs/2110.09522}{arXiv:2110.09522}{]}
that the expression for the Sagnac coordinate time delay given in
{[}\href{https://arxiv.org/abs/2109.14641}{arXiv:2109.14641}{]} ``differs
from the standard interpretation described in the book by Landau-Lifshitz
(LL)''. We note that: 1) the Sagnac effect is not even discussed
in LL; 2) the expression in {[}\href{https://arxiv.org/abs/2109.14641}{arXiv:2109.14641}{]}
is standard, given in countless papers and even textbooks; 3) the
expression by LL quoted by the author consists of the (infinitesimal)
two-way trip travel time for a light signal, which the author confuses
with the Sagnac time delay (when they are actually very different
things); 4) such confusion would negate the existence, both in special
and general relativity, of the well-known and experimentally tested
Sagnac effect; 5) the claims that it sheds doubt in any of the assertions
made in {[}\href{https://arxiv.org/abs/2109.14641}{arXiv:2109.14641}{]}
are completely unfounded. 
\end{abstract}

\section{Two-way trip of light signals, and Landau-Lifshitz space metric }

The line element $ds^{2}=g_{\alpha\beta}dx^{\alpha}dx^{\beta}$ of
a stationary spacetime can generically be written as 
\begin{equation}
ds^{2}=-e^{2\Phi}(dt-\mathcal{A}_{i}dx^{i})^{2}+h_{ij}dx^{i}dx^{j}\ ,\label{eq:StatMetric}
\end{equation}
where $e^{2\Phi}=-g_{00}$, $\Phi\equiv\Phi(x^{j})$, $\mathcal{A}_{i}\equiv\mathcal{A}_{i}(x^{j})=-g_{0i}/g_{00}$,
and $h_{ij}\equiv h_{ij}(x^{k})=g_{ij}+e^{2\Phi}\mathcal{A}_{i}\mathcal{A}_{j}$.

Consider two infinitesimally close observers at rest in the coordinates
of \eqref{eq:StatMetric}: observer $E$, carrying a flashlight, and
observer $R$, carrying a mirror. Observer $E$ emits a light flash
at position $x_{E}^{i}$, which is reflected by observer $R$'s mirror
at $x_{R}^{i}=x_{E}^{i}+dx^{i}$, returning then to $E$; see Fig.~18
in Sec.~\S84 of the Landau-Lifshitz (LL) textbook \cite{LandauLifshitz}.
Along the photon's worldline, $ds^{2}=0$; by (\ref{eq:StatMetric}),
this yields two solutions for $dt$, of which the one corresponding
to a future-oriented null worldline is 
\begin{equation}
dt=\mathcal{A}_{i}dx^{i}+e^{-\Phi}dl\ ;\qquad dl=\sqrt{h_{ij}dx^{i}dx^{j}}\ \quad(\text{for a photon}).\label{eq:Photon}
\end{equation}
For the trips $E\rightarrow R$, and $R\rightarrow E$ we have, respectively,
\[
dt_{ER}=\mathcal{A}_{i}dx^{i}+e^{-\Phi}dl\ ;\qquad\quad dt_{RE}=-\mathcal{A}_{i}dx^{i}+e^{-\Phi}dl\ .
\]
Observe that the term $\mathcal{A}_{i}dx^{i}$, but not $dl$, changes
sign with an inversion of direction. Hence, for the photon's two-way
trip $E\rightarrow R\rightarrow E$, 
\begin{equation}
dt_{ER}+dt_{RE}=2e^{-\Phi}dl\ ,\label{eq:2way}
\end{equation}
which corresponds to the unnumbered equation below Eq. (84.5) in \cite{LandauLifshitz}
quoted by the author of \cite{Deriglazov}.\footnote{LL write this as the {\em difference} between the coordinate time
intervals calculated with respect to the reflection event (thus one
being typically positive and the other typically negative); this is
perhaps the source of the confusion by the author of \cite{Deriglazov}.} In terms of observer $E$'s proper time, $d\tau=e^{\Phi}dt$, this
time interval equals twice $dl$, which is thus the measured spatial
distance between $E$ and $R$. This standardly defines $h_{ij}$
as the spatial metric, introduced in \cite{LandauLifshitz}, and subsequently
widely used in the literature on 1+3 spacetime splittings (e.g. \cite{ZonozBell1998,ManyFaces,NatarioQM2007,Analogies,Zonoz2019,Cilindros,MinguzziSimultaneity},
including \cite{PaperDragging}).

From the above discussion it follows that integrating the quantity
\begin{equation}
dl=e^{\Phi}(dt-\mathcal{A}_{i}dx^{i})\equiv dt_{p}\label{dl}
\end{equation}
along the worldline of a photon propagating in an optical fiber, as
suggested by the author of \cite{Deriglazov} (but \emph{not} by LL),
will simply result in the length of the optical fiber as measured
by the observers at rest in the coordinates of \eqref{eq:StatMetric}
(which of course is the same in both directions), see Fig.~\ref{fig:Spacetime-diagram}(b).
Notice that such integral does not even represent the time measured
by some observer $E$ for a finite photon two-way trip along the optical
fiber (unless $\Phi$ is constant along it), since such interval would,
by Eq. \eqref{eq:Photon}, be $\Delta\tau_{ERE}=e^{\Phi}(\Delta t_{ER}+\Delta t_{RE})=2e^{\Phi}\int_{E}^{R}e^{-\Phi}dl$.
More importantly: \emph{\uline{this has nothing to do with the
Sagnac effect}}.

\section{Sagnac effect}

\begin{figure}
\includegraphics[width=0.6\paperwidth]{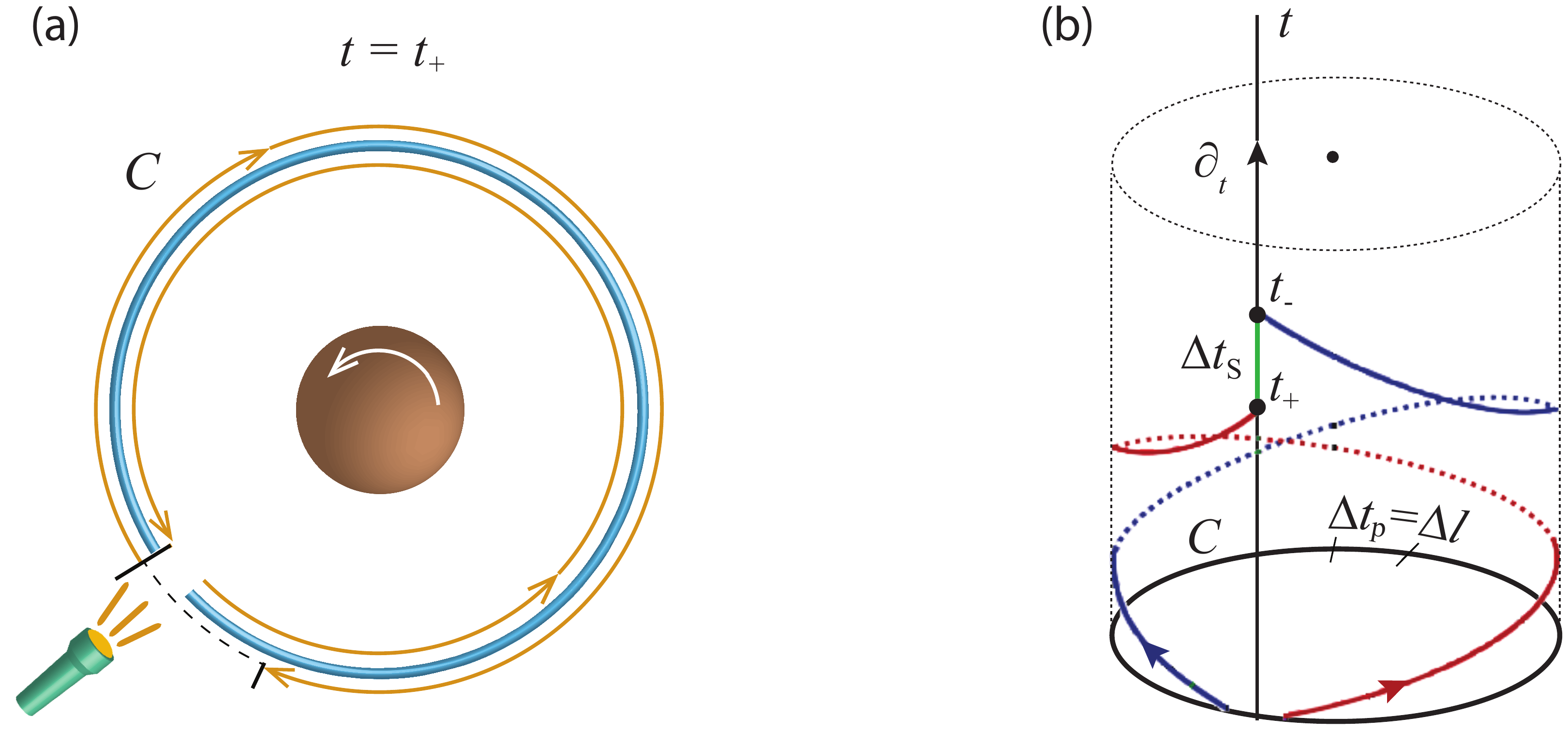}\caption{\label{fig:Spacetime-diagram}(a) Light beams propagating in opposite
directions along an optical fiber loop $C$ around a spinning body
(Fig. 1(a) of \cite{PaperDragging}). (b) Spacetime diagram for this
setup (inspired on Fig.~2 of \cite{Kajari:2009qy}): the difference
in arrival times $\Delta t_{{\rm S}}$ in Eq. \eqref{eq:SagnacDiffForm}
(Sagnac \emph{coordinate} time delay) is an interval along a $t$-coordinate
line (which translates, in the observer's proper time, to $\Delta\tau_{{\rm S}}=e^{\Phi}\Delta t_{{\rm S}}$).
The quantity $t_{p}=\int_{C}dl$, dubbed \textquotedblleft true time\textquotedblright{}
in \cite{Deriglazov}, yields in fact the length of the spatial curve
$C$. Confusing it with the beams' travel time would lead to the erroneous
conclusion that they arrive at the same time, in contradiction with
measurement (by e.g. a Sagnac interferometer).}
\end{figure}

The Sagnac effect (e.g. \cite{SagnacI,SagnacII,Laue1920,Post1967,AshtekarMagnon,Chow_et_al1985,CiufoliniWheeler,Tartaglia:1998rh,Kajari:2009qy,BiniJantzenMashhoon_Clock1,Kajari:2004ms,MinguzziSimultaneity,AshbyLivinh,AshbyBook,SoffelBook1989,RizziRuggieroAharonovII,Gourgoulhon:SpecialRelativ,Ruggiero_SagnacTestsReview,Tartaglia-et-al-RingLasers2016,TartagliaExperimental2020,Bosi:2011um,Rincon:2019zxk}%
) consists of the difference in arrival times of light beams propagating
in opposite directions around a spatially closed loop $C$. Consider
an optical fiber loop, where observer $E$ {[}at rest in the coordinates
of \eqref{eq:StatMetric}{]} injects light beams in opposite directions,
as depicted in Fig.~\ref{fig:Spacetime-diagram} (a) (similar to
Fig. 1 (a) of \cite{PaperDragging}). Using the $+$ ($-$) sign to
denote the anti-clockwise (clockwise) directions, the coordinate time
it takes for a full loop is, from Eq. \eqref{eq:Photon}, respectively

\[
t_{\pm}=\oint_{\pm C}dt=\oint_{C}e^{-\Phi}dl\pm\oint_{C}\mathcal{A}_{i}dx^{i}\ ;
\]
therefore, the Sagnac time delay is, in coordinate time and in observer
$E$'s proper time, respectively (cf.~Eq.~(4) of~\cite{PaperDragging}),
\begin{equation}
\Delta t_{{\rm S}}\equiv t_{+}-t_{-}=2\oint_{C}\mathcal{A}_{i}dx^{i}=2\oint_{C}\bm{\mathcal{A}}\ ,\qquad\quad\ \quad\Delta\tau_{{\rm S}}=e^{\Phi}\Delta t_{{\rm S}}\label{eq:SagnacDiffForm}
\end{equation}
(where we identified $\mathcal{A}_{i}dx^{i}$ with the spatial 1-form
$\bm{\mathcal{A}}\equiv\mathcal{A}_{i}\mathbf{d}x^{i}$). Equation
\eqref{eq:SagnacDiffForm} is a well known, \emph{standard} result,
given, \emph{precisely in (one, or both of) these two forms}, in e.g.
\cite{AshtekarMagnon,Kajari:2009qy,Kajari:2004ms,MinguzziSimultaneity,BiniJantzenMashhoon_Clock1,Cilindros,Ruggiero_SagnacTestsReview,Tartaglia-et-al-RingLasers2016,TartagliaExperimental2020,Bosi:2011um,Rincon:2019zxk,CiufoliniWheeler}.
It applies both to rotating frames in flat spacetime as well as to
arbitrary stationary gravitational fields, has been thoroughly experimentally
tested%
, plays a key role in the relativistic corrections for the Global
Position System \cite{AshbyLivinh,AshbyBook}, and is the basis of
optical gyroscopes \cite{Chow_et_al1985,Post1967,CiufoliniWheeler,WikipediaRingLaser,WikipediaOptical,GyroscopesReviews,CommercialOpticalGyros},
used as a means of detecting the absolute rotation of an apparatus
with respect to inertial frames, with broad commercial applications
nowadays \cite{CiufoliniWheeler,Gourgoulhon:SpecialRelativ,GyroscopesReviews,CommercialOpticalGyros,WikipediaRingLaser,WikipediaOptical}.
Based on this equation, proposals for experimental detection of frame-dragging
have also been put forth \cite{Chow_et_al1985,CiufoliniWheeler,TartagliaExperimental2020,Ruggiero_SagnacTestsReview,Tartaglia-et-al-RingLasers2016,Bosi:2011um}.

Observe the crucial difference between \eqref{eq:2way} and \eqref{eq:SagnacDiffForm}:
the former \emph{\uline{adds}} travel times for trips on opposite
directions along the same spatial path, whereas the latter \emph{\uline{subtracts}}
them. This is the reason why only the even parity term $e^{-\Phi}dl$
contributes to \eqref{eq:2way}, while $\Delta t_{{\rm S}}$ results
of integrating the odd parity term $\mathcal{A}_{i}dx^{i}$.

Finally, we remark that although this effect is not explicitly discussed
in LL \cite{LandauLifshitz}, a closely related quantity --- the
synchronization gap along $C$, which is well known (e.g. \cite{BiniJantzenMashhoon_Clock1,MinguzziSimultaneity})
to be one half the corresponding Sagnac time delay --- is computed
therein, see Eq. (88.5) in Sec. \S88; the result is, as would be
expected, \emph{exactly} $\Delta t_{{\rm S}}/2$, as given by Eq.
\eqref{eq:SagnacDiffForm} above (Eq.~(4) in~\cite{PaperDragging}).

\section{Conclusions}

To conclude, we have shown that the claims in \cite{Deriglazov} regarding
the results presented in \cite{PaperDragging} are completely unfounded:
the Sagnac effect is not even discussed in \cite{LandauLifshitz}
(just the closely related synchronization gap, and with results \emph{entirely}
consistent with those in \cite{PaperDragging}); the integral (5)
in \cite{Deriglazov}, that the author dubs ``true time'', does
not actually correspond to the travel time as measured by any observer,
but instead to the usual definition of spatial length; and the expression
for the Sagnac coordinate time delay in \cite{PaperDragging} is completely
standard.

\bibliographystyle{../utphys}
\bibliography{Ref}

\providecommand{\href}[2]{#2}\begingroup\raggedright\begin{thebibliography}{10}

\bibitem{LandauLifshitz}
L.~D. Landau and E.~M. Lifshitz, {\em {The classical theory of fields; 4rd
  ed.}}, vol.~2 of {\em Course of theoretical physics}.
\newblock {Butterworth-Heinemann}, Oxford, UK, 1975.
\newblock Trans. from the Russian.

\bibitem{Deriglazov}
A.~A. Deriglazov, ``{Comment on ''Frame-dragging: meaning, myths, and
  misconceptions'' by L. F. O. Costa and J. Nat\'ario},''
  \href{http://arxiv.org/abs/2110.09522}{{\ttfamily arXiv:2110.09522 [gr-qc]}}.

\bibitem{ZonozBell1998}
D.~Lynden-Bell and M.~Nouri-Zonoz, ``Classical monopoles: Newton, nut space,
  gravomagnetic lensing, and atomic spectra,''
  \href{http://dx.doi.org/10.1103/RevModPhys.70.427}{{\em Rev. Mod. Phys.}
  {\bfseries 70} (Apr, 1998) 427--445}.

\bibitem{ManyFaces}
R.~T. Jantzen, P.~Carini, and D.~Bini, ``{The Many faces of
  gravitoelectromagnetism},''
  \href{http://dx.doi.org/10.1016/0003-4916(92)90297-Y}{{\em Annals Phys.}
  {\bfseries 215} (1992) 1--50},
  \href{http://arxiv.org/abs/gr-qc/0106043}{{\ttfamily arXiv:gr-qc/0106043}}.

\bibitem{NatarioQM2007}
J.~Nat{\'a}rio, ``{Quasi-Maxwell interpretation of the spin--curvature
  coupling},'' \href{http://dx.doi.org/10.1007/s10714-007-0474-7}{{\em General
  Relativity and Gravitation} {\bfseries 39} no.~9, (Sep, 2007) 1477--1487}.

\bibitem{Analogies}
L.~F.~O. Costa and J.~Nat{\'a}rio, ``Gravito-electromagnetic analogies,''
  \href{http://dx.doi.org/10.1007/s10714-014-1792-1}{{\em General Relativity
  and Gravitation} {\bfseries 46} no.~10, (2014) 1792},
  \href{http://arxiv.org/abs/1207.0465}{{\ttfamily arXiv:1207.0465}}.

\bibitem{Zonoz2019}
R.~Gharechahi, J.~Koohbor, and M.~Nouri-Zonoz, ``{General relativistic analogs
  of Poisson's equation and gravitational binding energy},''
  \href{http://dx.doi.org/10.1103/PhysRevD.99.084046}{{\em Phys. Rev. D}
  {\bfseries 99} (2019) 084046}.

\bibitem{Cilindros}
L.~F.~O. Costa, J.~Nat\'ario, and N.~O. Santos, ``{Gravitomagnetism in the
  Lewis cylindrical metrics},''
  \href{http://dx.doi.org/10.1088/1361-6382/abc570}{{\em Class. Quant. Grav.}
  {\bfseries 38} no.~5, (2021) 055003},
  \href{http://arxiv.org/abs/1912.09407}{{\ttfamily arXiv:1912.09407}}.

\bibitem{MinguzziSimultaneity}
E.~Minguzzi, ``{Simultaneity and generalized connections in general
  relativity},'' \href{http://dx.doi.org/10.1088/0264-9381/20/11/332}{{\em
  Class. Quant. Grav.} {\bfseries 20} (2003) 2443--2456},
  \href{http://arxiv.org/abs/gr-qc/0204063}{{\ttfamily arXiv:gr-qc/0204063}}.

\bibitem{PaperDragging}
L.~F.~O. Costa and J.~Nat\'ario, ``{Frame-Dragging: Meaning, Myths, and
  Misconceptions},'' \href{http://dx.doi.org/10.3390/universe7100388}{{\em
  Universe} {\bfseries 7} no.~10, (2021) 388},
  \href{http://arxiv.org/abs/arXiv:2109.14641}{{\ttfamily
  arXiv:arXiv:2109.14641}}.

\bibitem{SagnacI}
G.~Sagnac, ``{L'\'{e}ther lumineux d\'{e}montr\'{e} par l'effet du vent relatif
  d'\'{e}ther dans un interf\'{e}rom\`{e}tre en rotation uniforme},'' {\em C.
  R. Acad. Sci., Paris} {\bfseries 157} (1913) 708--710.

\bibitem{SagnacII}
G.~Sagnac, ``{Sur la preuve de la r\'{e}alit\'{e} de l'\'{e}ther lumineux par
  l'exp\'{e}rience de l'interf\'{e}rographe tournant},'' {\em C. R. Acad. Sci.,
  Paris} {\bfseries 157} (1913) 1410--1413.

\bibitem{Laue1920}
M.~v.~Laue, ``{Zum Versuch von F. Harress},''
  \href{http://dx.doi.org/10.1002/andp.19203671303}{{\em Annalen der Physik}
  {\bfseries 367} no.~13, (1920) 448--463}.

\bibitem{Post1967}
E.~J. Post, ``Sagnac effect,''
  \href{http://dx.doi.org/10.1103/RevModPhys.39.475}{{\em Rev. Mod. Phys.}
  {\bfseries 39} (Apr, 1967) 475--493}.

\bibitem{AshtekarMagnon}
A.~Ashtekar and A.~Magnon, ``{The Sagnac effect in general relativity},''
\href{http://dx.doi.org/10.1063/1.522521}{{\em J. Math. Phys.} {\bfseries 16}
  (1975) 341--344}.

\bibitem{Chow_et_al1985}
W.~W. Chow {\em et~al.}, ``The ring laser gyro,''
  \href{http://dx.doi.org/10.1103/RevModPhys.57.61}{{\em Rev. Mod. Phys.}
  {\bfseries 57} (Jan, 1985) 61--104}.

\bibitem{CiufoliniWheeler}
I.~{Ciufolini} and J.~A. {Wheeler}, {\em {Gravitation and Inertia}}.
\newblock Princeton Series in Physics, Princeton, NJ, 1995.

\bibitem{Tartaglia:1998rh}
A.~Tartaglia, ``{General relativistic corrections to the Sagnac effect},''
  \href{http://dx.doi.org/10.1103/PhysRevD.58.064009}{{\em Phys. Rev. D}
  {\bfseries 58} (1998) 064009},
  \href{http://arxiv.org/abs/gr-qc/9806019}{{\ttfamily arXiv:gr-qc/9806019}}.

\bibitem{Kajari:2009qy}
E.~Kajari, M.~Buser, C.~Feiler, and W.~P. Schleich,
  \href{http://dx.doi.org/10.3254/978-1-58603-990-5-45}{``{Rotation in
  relativity and the propagation of light},''} in {\em {Proceedings of the
  International School of Physics "Enrico Fermi", Course CLXVIII}}, pp.~45--148
  [Riv. Nuovo Cim. \textbf{32}, 339--438].
\newblock IOS Press, Amsterdam, {2009}.
\newblock \href{http://arxiv.org/abs/0905.0765}{{\ttfamily arXiv:0905.0765}}.

\bibitem{BiniJantzenMashhoon_Clock1}
D.~Bini, R.~T. Jantzen, and B.~Mashhoon, ``{Gravitomagnetism and relative
  observer clock effects},''
  \href{http://dx.doi.org/10.1088/0264-9381/18/4/306}{{\em Class. Quant. Grav.}
  {\bfseries 18} (2001) 653--670},
  \href{http://arxiv.org/abs/gr-qc/0012065}{{\ttfamily arXiv:gr-qc/0012065}}.

\bibitem{Kajari:2004ms}
E.~Kajari, R.~Walser, W.~P. Schleich, and A.~Delgado, ``{Sagnac effect of
  Godel's universe},''
  \href{http://dx.doi.org/10.1023/B:GERG.0000046184.03333.9f}{{\em Gen. Rel.
  Grav.} {\bfseries 36} (2004) 2289},
  \href{http://arxiv.org/abs/gr-qc/0404032}{{\ttfamily arXiv:gr-qc/0404032}}.

\bibitem{AshbyLivinh}
N.~{Ashby}, ``{Relativity in the Global Positioning System},''
  \href{http://dx.doi.org/10.12942/lrr-2003-1}{{\em Living Reviews in
  Relativity} {\bfseries 6} no.~1, (Jan., 2003) 1}.

\bibitem{AshbyBook}
N.~Ashby, \href{http://dx.doi.org/10.1007/978-94-017-0528-8_3}{``The sagnac
  effect in the global positioning system,''} in {\em Relativity in Rotating
  Frames.}, R.~M.~e. Rizzi~G., ed., pp.~215--258 [Fund. Theor. Phys.
  \textbf{135}, 215].
\newblock Springer, Dordrecht, {2004}.

\bibitem{SoffelBook1989}
M.~H. {Soffel}, \href{http://dx.doi.org/10.1007/978-3-642-73406-9}{{\em
  {Relativity in Astrometry, Celestial Mechanics and Geodesy}}}.
\newblock Springer-Verlag, Berlin, Heidelberg, 1989.

\bibitem{RizziRuggieroAharonovII}
G.~Rizzi and M.~L. Ruggiero,
  \href{http://dx.doi.org/10.1007/978-94-017-0528-8}{``{The Relativistic Sagnac
  effect: Two derivations},''} in {\em {Relativity in Rotating Frames}},
  G.~Rizzi and M.~L. Ruggiero, eds., pp.~179--220.
\newblock Kluwer Academic Publishers, Dordrecht, {2004}.
\newblock \href{http://arxiv.org/abs/gr-qc/0305084}{{\ttfamily
  arXiv:gr-qc/0305084}}.

\bibitem{Gourgoulhon:SpecialRelativ}
E.~Gourgoulhon, \href{http://dx.doi.org/10.1007/978-3-642-37276-6}{{\em
  {Special Relativity in General Frames}}}.
\newblock Graduate Texts in Physics. Springer, Berlin, Heidelberg, 2013.

\bibitem{Ruggiero_SagnacTestsReview}
M.~L. Ruggiero, ``{Sagnac Effect, Ring Lasers and Terrestrial Tests of
  Gravity},'' \href{http://dx.doi.org/10.3390/galaxies3020084}{{\em Galaxies}
  {\bfseries 3} no.~2, (2015) 84--102},
  \href{http://arxiv.org/abs/1505.01268}{{\ttfamily arXiv:1505.01268}}.

\bibitem{Tartaglia-et-al-RingLasers2016}
A.~Tartaglia, A.~Di~Virgilio, J.~Belfi, N.~Beverini, and M.~L. Ruggiero,
  ``{Testing general relativity by means of ring lasers},''
  \href{http://dx.doi.org/10.1140/epjp/i2017-11372-5}{{\em Eur. Phys. J. Plus}
  {\bfseries 132} no.~2, (2017) 73},
  \href{http://arxiv.org/abs/1612.09099}{{\ttfamily arXiv:1612.09099}}.

\bibitem{TartagliaExperimental2020}
A.~Tartaglia, ``{Relativistic positioning and Sagnac-like measurements for
  fundamental physics in space},''
  \href{http://dx.doi.org/10.1016/j.asr.2020.05.039}{{\em Adv. Space Res.}
  {\bfseries 66} (2020) 2757--2763},
  \href{http://arxiv.org/abs/2005.13397}{{\ttfamily arXiv:2005.13397}}.

\bibitem{Bosi:2011um}
F.~Bosi {\em et~al.}, ``{Measuring Gravito-magnetic Effects by Multi Ring-Laser
  Gyroscope},'' \href{http://dx.doi.org/10.1103/PhysRevD.84.122002}{{\em Phys.
  Rev. D} {\bfseries 84} (2011) 122002},
  \href{http://arxiv.org/abs/1106.5072}{{\ttfamily arXiv:1106.5072}}.

\bibitem{Rincon:2019zxk}
A.~Rinc\'on and J.~R. Villanueva, ``{The Sagnac effect on a scale-dependent
  rotating BTZ black hole background},''
  \href{http://dx.doi.org/10.1088/1361-6382/aba17f}{{\em Class. Quant. Grav.}
  {\bfseries 37} no.~17, (2020) 175003},
  \href{http://arxiv.org/abs/1902.03704}{{\ttfamily arXiv:1902.03704}}.

\bibitem{WikipediaRingLaser}
``Ring laser gyroscope.''
\newblock \url{https://en.wikipedia.org/wiki/Ring_laser_gyroscope}.

\bibitem{WikipediaOptical}
``Fibre-optic gyroscope.''
\newblock \url{https://en.wikipedia.org/wiki/Fibre-optic_gyroscope}.

\bibitem{GyroscopesReviews}
V.~Passaro {\em et~al.}, ``Gyroscope technology and applications: A review in
  the industrial perspective,'' \href{http://dx.doi.org/10.3390/s17102284}{{\em
  Sensors} {\bfseries 17} no.~10, (2017) }.

\bibitem{CommercialOpticalGyros}
H.~Kajioka {\em et~al.},
  \href{http://dx.doi.org/10.1117/12.258177}{``{Commercial applications of
  mass-produced fiber optic gyros},''} in {\em Fiber Optic Gyros: 20th
  Anniversary Conference}, E.~Udd, H.~C. Lefevre, and K.~Hotate, eds.,
  vol.~2837, pp.~18 -- 32, International Society for Optics and Photonics.
\newblock SPIE, 1996.

\end{thebibliography}\endgroup

\end{document}